\documentclass[10pt]{article}  
\usepackage{graphicx,floatflt,amssymb,epsf}   
\def\fa{${\bf \bar{5}_1}$}  
\def\fb{${\bf \bar{5}_2}$}  
\def\fc{${\bf \bar{5}_3}$}  
\def\ffi{${\bf \bar{5}_i}$}  
\def\fj{${\bf \bar{5}_j}$}  
 
\def\ta{${\bf {10}_1}$}  
\def\tb{${\bf {10}_2}$}  
\def\tc{${\bf {10}_3}$} 
\def\ti{${\bf {10}_i}$}

\def\ni{${\bf {N}_i}$}

\def\smg{${\rm SU(3)}_{\rm C}\otimes {\rm SU(2)}_{\rm L}  
\otimes {\rm U(1)}_{\rm Y}$}  
  
\def\e{\epsilon} 
 
\begin{document}  
\begin{flushright}  
\texttt{hep-ph/0511276}\\  
SINP/TNP/05-27\\  
%CU-PHYSICS/??-2005\\ 
\end{flushright}  
  
\vskip 3pt  
  
\begin{center}  
{\Large \bf Remarks on flavour mixings from orbifold compactification} \\  
\vspace*{1cm}  
\renewcommand{\thefootnote}{\fnsymbol{footnote}}  
{\large {\sf Gautam Bhattacharyya${}^1$}   
and {\sf Amitava Raychaudhuri${}^2$}  
} \\  
\vspace{5pt}  
{\small ${}^{1)}$ {\em Saha Institute of Nuclear Physics,  
1/AF Bidhan Nagar, Kolkata 700064, India} \\   
   ${}^{2)}$ {\em Department of Physics, University of Calcutta,  
92 A.P.C. Road, Kolkata 700009, India}}   
  
\normalsize  
\end{center}  
  
\begin{abstract}  
  We consider 5d SU(5) GUT models based on the orbifold $S^1/(Z_2 
  \times Z_2')$, and study the different possibilities of placing  
  the SU(5) matter multiplets in three possible locations, namely, 
  the two branes at the two orbifold fixed points and SU(5) bulk. We 
  demonstrate that if flavour hierarchies originate solely from 
  geometrical suppressions due to wavefunction normalisation of fields 
  propagating in the bulk, then it is not possible to satisfy even the 
  gross qualitative behaviour of the CKM and MNS matrices regardless 
  of where we place the matter multiplets.

\vskip 3pt \noindent  
\texttt{PACS Nos:~11.10.Kk, 11.30.Hv} \\ 
\texttt{Key Words:~Flavour, Orbifold compactification}  
\end{abstract}

\renewcommand{\thesection}{\Roman{section}}  
\setcounter{footnote}{0}  
\renewcommand{\thefootnote}{\arabic{footnote}}

\section{Introduction} 
In order to bypass some of the problems with conventional Grand Unified
Theories (GUTs), namely, the doublet-triplet splitting, too fast proton decay,
etc, new proposals have been advanced which allow realization of GUT gauge
symmetry in a higher dimensional orbifold \cite{hmos}-\cite{bs}. The
quantitative success of supersymmetric (SUSY) gauge coupling unification
provides further attraction to embed SUSY in a higher dimensional SU(5) gauge
symmetry. Several attempts have been made in this regard. In this short note,
we consider those which have only one extra dimension compactified on the
orbifold $S^1/(Z_2 \times Z_2')$. We go by the hypothesis that flavour
hierarchies originate {\em purely} from geometrical suppression of Yukawa
couplings depending on the location of the fields attached to the given vertex
in the extra dimension. There are three possible locations where matter fields
can be placed: the SU(5) preserving five-dimensional (5d) bulk, and the two
fixed points, namely, the SU(5) preserving $O$ brane ($y = 0$), and the
standard model (SM) brane $O'$ ($y = \pi R/2$), where $y$ is the 5th
coordinate and $R$ is the radius of compactification. If a particular field
extends in the bulk, it is associated with a wavefunction normalisation factor
($< 1$), whereas for a field restricted to one of the two branes there is no
such suppression. So the relative size of two Yukawa couplings comes from the
relative number of brane vis-a-vis bulk fields attached to them.  In this
brief communication, we investigate whether by placing the SU(5) multiplets of
different generations (\ffi, \ti), including the right-handed singlet
neutrinos (\ni), in different locations, one can generate even the overall
qualitative characteristics of the Cabibbo-Kobayashi-Maskawa (CKM) and
Maki-Nakagawa-Sakata (MNS) matrices: the question of quantitative success
comes thereafter.  We examine all possible choices for the relative placements
of the multiplets of different generations at alternate locations in a
systematic manner and demand that {\em only} locations matter in determining
the relative Yukawa couplings. We observe that each such case fails to meet
this test.  One can, however, argue that O(1) corrections to the Yukawa
couplings are in general unavoidable. Such corrections, as mentioned in
\cite{hmos}-\cite{cn}, play a key r\^{o}le in generating flavor
mixings. Clearly, this is achieved at the cost of introducing many new and
unknown parameters. In this short note, we reach at a similar conclusion by
looking at the problem from a different angle.  We show on a case by case
basis that if hierarchies emerge purely from geometrical suppressions, barring
any further O(1) corrections for economy of parameters, it is not
possible to simultaneously satisfy even the gross features of the three
generation CKM and MNS mixings, no matter where we place the different
multiplets.

\section{Formalism}  
We briefly summarise the primary consequences of orbifold compactification. 
Let us conceive a 5d GUT with minimal SU(5) gauge group and $N = 1$ SUSY. The 
5d spacetime is factorised into a product of the 4d spacetime $M^4$ (labelled 
by the coordinates $x^{\mu}$) with the extra spatial dimension compactified on 
the orbifold $S^1/(Z_2 \times Z_2')$ (labelled by the coordinate 
$y=x_5$). The inverse radius $R^{-1}$ is chosen to be of the order of $M_{\rm 
GUT} = 10^{16}$ GeV.  The orbifold construction proceeds by dividing $S^1$ 
first by a $Z_2$ transformation $y \rightarrow -y$ and then by a further 
division by $Z_2'$ which acts as $y' \rightarrow -y'$ with $y'=y+\pi 
R/2$. After these identifications, the physical spacetime becomes the interval 
$[0, \pi R/2]$ with a brane located at each fixed point $y=0$ and $y=\pi 
R/2$. As a result of the two reflections, the branes at $y=\pi R$ and $-\pi 
R/2$ are identified with those at $y=0$ and $y=\pi R/2$, respectively.  Now 
let us consider a generic field $\phi(x^{\mu}, y)$ existing in the 5d 
bulk. The $Z_2$ and $Z_2'$ parities (called $P$ and $P'$, respectively) are 
defined for this field as 
\begin{eqnarray} 
\phi(x^{\mu},y) &\rightarrow & \phi(x^{\mu},-y) = P\phi(x^{\mu},y) , 
\nonumber \\ 
\phi(x^{\mu},y') &\rightarrow & \phi(x^{\mu},-y') = P'\phi(x^{\mu},y') . 
\end{eqnarray} 
Using the notation $\phi_{\pm\pm}$ for the fields with $(P,P')=(\pm, \pm)$, we
are led to the following observations regarding the 4d KK fields.
$\phi_{++}^{(2n)}$ acquire a mass $2n/R$, while $\phi_{+-}^{(2n+1)}$ and
$\phi_{-+}^{(2n+1)}$ acquire a mass $(2n+1)/R$ and $\phi_{--}^{(2n+2)}$
acquire a mass $(2n+2)/R$. This implies that the only fields which can have
massless components are $\phi_{++}^{(2n)}$. The other interesting consequence
is that only $\phi_{++}$ and $\phi_{+-}$ can have non-vanishing components on
the $y=0$ brane. In fact, compactification leads to symmetry reduction.  The
starting theory is 5d $N = 1$ SUSY invariant under the gauge group SU(5). From
a 4d perspective, this is equivalent to $N = 2$ SUSY. We assign 
suitable $(P,P')$ quantum numbers
for the fields. Upon the first compactification
by $Z_2$ the conjugated fields are projected out and the $N = 2$ SUSY reduces
to $N = 1$ SUSY but still respecting the gauge SU(5); on the second
compactification by $Z_2'$ the SU(5) gauge symmetry is broken to the SM gauge
group \smg~ with an unbroken $N = 1$ SUSY\footnote{The doublet-triplet
splitting problem is elegantly solved as the coloured triplet Higgs do not
have $(++)$ assignments unlike the Higgs doublets, as a result the former has
a mass of order $1/R \sim M_{\rm GUT}$.}.

The $P$ and $P'$ quantum numbers are to be so arranged that the 5d SU(5) gauge 
symmetry remains intact at $O$ but is broken to \smg~ at $O'$. This can be 
done by choosing $P = (+++++)$ and $P'= (---++)$ or $(+++--)$ acting on {\bf 
5}. As has been noted \cite{hmos,hm1}, with the above $P'$ assignments it is 
not possible to fill up a complete SU(5) multiplet by zero mode matter. One 
has to introduce ${\bf \bar{5}'}$ and ${\bf 10'}$ with $P'$ assignments 
opposite to those in ${\bf \bar{5}}$ and ${\bf 10}$ to obtain correct low 
energy matter content\footnote{If \fa and \ta are kept in SU(5) bulk, then the 
first generation zero mode quarks and leptons come from different SU(5) 
multiplets, and proton decay from broken gauge boson exchange does not exist 
at leading order.}. 
 
Now we come to the discussion of Yukawa couplings. All such couplings 
consistent with gauge symmetry and R-parity are admitted. We assume 
that the hierarchical structure of the effective 4d Yukawa couplings 
is generated solely from the different normalisation of brane and bulk 
fields. Let us denote\footnote{See, e.g. the Lagrangian in Eq.~(6) of 
  \cite{hmos}.} a Yukawa coupling involving three brane superfields by 
$\lambda$. The Yukawa coupling for an interaction where one of the 
three fields is replaced by the zero mode of a bulk field is $ 
\lambda/\sqrt{M_* R}$, where $M_*$ is the UV cutoff scale of the 5d 
theory and the appearance of $M_*$ is related to the canonical 
normalisation of the zero mode kinetic terms\footnote{$M_* R \sim 
10^{2-3}$ is a good choice for gauge coupling unification 
\cite{nomura}.}. The number of $\e = 1/\sqrt{M_* R}$ factors in 
front of $\lambda$ is given by the number of bulk zero modes in a 
given interaction, each bulk field contributing one such factor.

\section{Fermion mass matrices}  
We write the fermion mass matrices in the convention that 
the fields on the left are left-handed and those on the right are 
right-handed.  The up quark mass matrix is given by $\overline{\bf 
  10}_i (M_u)_{ij} {\bf 10^c}_j$ and is hermitian.  The down quark mass 
matrix is given by $\overline{\bf 10}_i (M_d)_{ij} {\bf 5^c}_j$.  The 
charged lepton mass matrix $M_l$ is simply $(M_d)^\dagger$.  For 
illustration, $M_l$ can be schematically represented as: 
\begin{equation} 
\begin{array}{cc}  
{\begin{array}{ccc} 
\;\;\;\;\;\;\; \; $\fa$ \; \; &  $\fb$ & 
\; \;$\fc$\\ 
\end{array}} 
& \\ 
M_l = \left\{ 
{\begin{array}{ccc} 
a \;\;&\;\;b \;\;& c \\ 
d \;\;&\;\;e \;\;& \;\;f \\ 
g \;\;&\;\;h \;\;& \;\; i \\ 
\end{array}} 
\right\} 
& 
{\begin{array}{c} 
\!\!\!{\bf {10}_1^c} \\ 
\!\!\!{\bf {10}_2^c} \\ 
\!\!\!{\bf {10}_3^c} \\ 
\end{array}} \\ 
\end{array}\;. 
\end{equation} 
where the entries $a,b, \ldots, i$ are determined solely by the 
locations of the \ti ~and \ffi ~fields. 
 
The neutrino mass\footnote{One can also generate Majorana mass matrix of the
  form $\bar{\bf 5}_i (M_M)_{ij} \bar{\bf 5}_j$ for light neutrinos via
  see-saw mechanism by integrating out the heavy ${\bf N}_i$ fields. Even then
  our conclusions below will go through.}
 is given by $\bar{\bf 5}_i (M_\nu)_{ij} {\bf N}_j$. 
In general, $M_d$, $M_l$ and $M_\nu$ are not hermitian, so they are 
diagonalised by biunitary transformations. The CKM matrix is given by 
$V_{\rm CKM} = V_u^\dagger V_d$, where $V_u$ diagonalises $M_u$ as 
$V_u^T M_u V_u = {\rm diag} (m_u, m_c, m_t)$. Similarly, $V_d$ 
diagonalises $M_d (M_d)^\dagger$. In the same way, the MNS matrix is 
given by $V_{\rm MNS} = V_\nu^\dagger V_l$, where $V_\nu$ and $V_l$ 
diagonalise $M_\nu (M_\nu)^\dagger$ and $M_l (M_l)^\dagger$ 
respectively\footnote{The relation $M_l = (M_d)^\dagger$ 
and the hermiticity of $M_u$ are valid in the SU(5) limit and 
will not be applicable when fermion 
multiplets are located on the SM brane $O'$. Our conclusions 
below are not affected by this.}.

\section{Results}\label{sec:res} 
In this section we consider one by one the possible cases 
distinguished by the locations of the different SU(5) matter 
multiplets and  examine whether the gross features of the CKM and 
MNS mixing matrices can be reproduced based on the geometric 
suppression factors alone. 
 
\begin{enumerate}    
\item \fa, \fb, \fc~ all at the same location is not allowed: In this 
  situation, as far as the contributions to $M_l$ are concerned, there 
  will be no difference between lepton flavours.  $M_l (M_l)^\dagger$ 
  will have a democratic structure because of the symmetry \ffi 
  $\leftrightarrow$ \fj ~for all $i,j =$ 1,2,3. $M_\nu 
  (M_\nu)^\dagger$ will also share the same structure. Consequently, 
  $V_l$ and $V_\nu$ will be identical and the MNS matrix will be the 
  identity matrix.  Note that it is only required that \fa, \fb, \fc 
  ~should not be at the same location; it does not matter whether this is 
  the SU(5) brane, the SM brane, or the bulk.

\item All \ti~ cannot be placed in the same location: This alternative 
  can be ruled out on grounds very similar to the previous one. 
  Because of the permutation symmetry between the \ti, now $M_u$ will 
  be a democratic matrix. This property will also be shared by $M_d 
  (M_d)^\dagger$. Thus, $V_u$ and $V_d$ will be identical and the 
  quark sector will remain unmixed. 
 
\item No two \ffi ~in the same location is allowed: Let \fa ~and \fb 
  ~share a location. In this case, the first and second rows and 
  columns of $M_l (M_l)^\dagger$ will be identical. This will also be 
  the case for $M_\nu (M_\nu)^\dagger$. Applying a unitary 
  transformation -- the same for both matrices -- they can be brought 
  to block diagonal forms with one state -- say `1' -- decoupled from 
  the other two -- `2' and `3'. This common unitary transformation 
  will have no impact on $V_{\rm MNS}$. Therefore, for this case, the 
  lepton sector mixing will be among two generations only. This 
  disagrees with the form determined by the data. 
 
\item Two \ti ~in the same location is not allowed: The argument in 
  this case is the same as that of the previous one excepting that one 
  now has to appeal to the matrices $M_u$ and $M_d (M_d)^\dagger$. Now 
  $V_{\rm CKM}$ will mix only two generations -- a situation 
  contradicting experimental requirements on flavour mixing. 
   
\item In view of the possibilities excluded above, the three \ti ~must 
  be in different locations and so should be the three \ffi. We now 
  show that placing (\ffi, \ti) pair-wise in the same location for 
  every $i$ is not allowed. As noted before, $M_u$ is a hermitian 
  matrix.  When \ffi ~and \ti ~are in the same location, then $M_u$ 
  and $M_d$ become proportional since according to our starting 
  hypothesis the entries in the matrix are determined by geometrical 
  considerations alone. Therefore, they are both diagonalised by the 
  same unitary transformation and mixing in the quark sector will 
  vanish. 
   
\item The only remaining possibility is that of having (\fj, \ti) 
  pair-wise placed at the same location for $i \neq j$. But even this 
  case is ruled out. This is because this alternative can be brought 
  to the form of the previous case by mere redefinitions of 
  rows/columns in $M_u$ and $M_d (M_d)^\dagger$.   
\end{enumerate}

\section{Discussions and Conclusions} 
In this brief note, we have considered a SUSY SU(5) theory defined on a 5d
space where the extra dimension is an $S^1/(Z_2 \times Z_2')$ orbifold. We
have used the hypothesis that the entries of the fermion mass matrices are
determined entirely by geometrical factors determined by the locations of the
SU(5) multiplets. We have shown that though there are many alternate
possibilities of locating the various fermion multiplets, in no case can one
reproduce even the qualitative nature of the CKM and MNS mixing matrices.
 
One way to get around this impasse is to invoke O(1) corrections to the
entries of the mass matrices. By this is meant that geometric factors only
determine the scale of an entry but its exact value is arbitrary. This would
plague the arguments we used above.  In such an event a qualitative or even a
quantitative success can be achieved.  But this is at the cost of a huge
arbitrariness since we have to acquiesce in a host of new parameters.
 
Some of the existing analyses have relied on the O(1) corrections to generate
nontrivial mixings. In \cite{hm2}, \ta~ has been kept at the $O'$ brane, \tb~
in the SU(5) bulk, while \tc~ has been placed at the $O$ brane.  All \ffi~
have been kept in the SU(5) bulk.  Going by our hypothesis, this case will
render the MNS mixing to be trivial {\em a la} our case (1) in section
\ref{sec:res}. In \cite{nomura}, the placements are the following: \fa~ and
\ta~ in bulk, \fb~ at $O$, \tb~ in bulk, \fc~ and \tc~ at $O$.  This
possibility contradicts reality as per our case (3) or (4).  In \cite{cn}, all
\ffi~ have been placed at $O$, while \ta~ and \tb~ reside in bulk with \tc~ at
$O$. Again, following our case (1), this option fails to reproduce observed
data.  The placement of matter fields in the above noted analyses have been
motivated from different considerations: suppressions of proton decay, $m_b =
m_\tau$ at GUT scale, etc.  We must admit one aspect at this stage. All these
analyses do emphasize the need of O(1) corrections to Yukawa couplings to
reproduce the data, i.e. the different entries of the mass matrices
constructed in these analyses should be taken merely as mass scales with the
tacit assumption that there are hidden O(1) uncertainty factors multiplying
those entries. In any case, our study is neither intended to nor does it in
any way undermine the different scenarios that the above mentioned analyses
deal with, rather we arrive at a similar conclusion.  Our modest intention is
to demonstrate by exhaustion that geometrical suppressions (depending on
localisation) alone, barring any O(1) corrections to Yukawa couplings, cannot
reproduce even the qualitative features of quark and lepton mixings in the 5d
SUSY GUT context. Any attempt to build a realistic model would be at the cost
of economy of parameters. Generalisation to 6d models with orbifolds of the
structure $T^2/(Z_2 \times Z_2')^2$ opens up more options for placing matter
fields in different locations \cite{hmos}. This allows further spatial
separation of these fields which help create textures that can admit
hierarchical masses with appropriate mixings. We do not deal with this in the
present analysis.

\vskip 10pt 
%{\bf Acknowledgements}:~
\centerline{\bf{Acknowledgements}}  
We acknowledge hospitality at Abdus Salam ICTP, Trieste, where part of 
the work has been done. This research has been supported, in part, by 
the DST, India, project number SP/S2/K-10/2001.


\begin{thebibliography}{99}   
 
\bibitem{hmos} 
  L.~Hall, J.~March-Russell, T.~Okui and D.~R.~Smith, 
  %``Towards a theory of flavor from orbifold GUTs,'' 
  JHEP {\bf 0409} (2004) 026 
  [arXiv:hep-ph/0108161]. 
 
 
 
\bibitem{hm1} 
  A.~Hebecker and J.~March-Russell, 
  %``A minimal S(1)/(Z(2) x Z'(2)) orbifold GUT,'' 
  Nucl.\ Phys.\ B {\bf 613} (2001) 3 
  [arXiv:hep-ph/0106166]. 
 
\bibitem{nomura} 
  Y.~Nomura, 
  %``Strongly coupled grand unification in higher dimensions,'' 
  Phys.\ Rev.\ D {\bf 65} (2002) 085036 
  [arXiv:hep-ph/0108170]. 
 
\bibitem{hm2} 
  A.~Hebecker and J.~March-Russell, 
  %``The flavour hierarchy and see-saw neutrinos from bulk masses in 5d 
  %orbifold GUTs,'' 
  Phys.\ Lett.\ B {\bf 541} (2002) 338 
  [arXiv:hep-ph/0205143]. 
 
\bibitem{hn} 
  L.~J.~Hall and Y.~Nomura, 
  %``Gauge unification in higher dimensions,'' 
  Phys.\ Rev.\ D {\bf 64} (2001) 055003 
  [arXiv:hep-ph/0103125]. 
 
\bibitem{cn} 
  W.~F.~Chang and J.~N.~Ng, 
  %``Neutrino masses in 5D orbifold SU(5) unification models without 
  %right-handed singlets,'' 
  JHEP {\bf 0310} (2003) 036 
  [arXiv:hep-ph/0308187]. 
 
\bibitem{af} 
  G.~Altarelli and F.~Feruglio, 
  %``SU(5) grand unification in extra dimensions and proton decay,'' 
  Phys.\ Lett.\ B {\bf 511} (2001) 257 
  [arXiv:hep-ph/0102301]. 
 
\bibitem{bs} 
  G.~Bhattacharyya and K.~Sridhar, 
  %``Testing orbifold models of supersymmetric grand unification,'' 
  J.\ Phys.\ G {\bf 29} (2003) 993 
  [arXiv:hep-ph/0111345]. 
 
 
\end{thebibliography}
\end{document}